# Rapid Development of Omics Data Analysis Applications through Vibe Coding

Jesse G. Meyer

Department of Computational Biomedicine, Cedars Sinai Medical Center, Los Angeles CA 90048

**Abstract**:

Building custom data analysis platforms traditionally requires extensive software engineering expertise, limiting accessibility for many researchers. Here, I demonstrate that modern large language models (LLMs) and autonomous coding agents can dramatically lower this barrier through a process called 'vibe coding'—an iterative, conversational style of software creation where users describe goals in natural language and AI agents generate, test, and refine executable code in real-time. As a proof of concept, I used Vibe coding to create a fully functional proteomics data analysis website capable of performing standard tasks, including data normalization, differential expression testing, and volcano plot visualization. The entire application, including user interface, backend logic, and data upload pipeline, was developed in less than ten minutes using only four natural-language prompts, without any manual coding, at a cost of under $2. Previous works in this area typically require tens of thousands of dollars in research effort from highly trained programmers. I detail the step-by-step generation process and evaluate the resulting code's functionality. This demonstration highlights how vibe coding enables domain experts to rapidly prototype sophisticated analytical tools, transforming the pace and accessibility of computational biology software development.

## Introduction

Mass-spectrometry–based proteomics depends on a multi-stage computational pipeline, beginning with raw data processing and peptide identification, followed by statistical modeling and interpretive analysis of protein abundance data. Over the past decade, a diverse ecosystem of tools has emerged to support these stages. At the raw-data level, MSFragger[1] and DIA-NN[2] have become among the most widely used frameworks for converting instrument files into quantitative protein matrices. These programs produce peptide- and protein-level quantification tables that serve as the input for the next stage of analysis—statistical modeling, visualization, and biological interpretation.

At this interpretive layer, Perseus remains one of the most influential tools[3]. Designed for users without programming experience, Perseus provides a graphical environment for normalization, imputation, clustering, enrichment analysis, and visualization, enabling biologists to perform complex analyses interactively. A new generation of web-based platforms has expanded on this accessibility by moving similar functionality into the browser. **ProteoArk** is a recent example, offering automated normalization, differential expression, and visualization pipelines that accept standard output formats from MaxQuant or Proteome Discoverer[4]. **MSstatsShiny** provides a web interface to the established MSstats statistical framework, simplifying quantitative analysis across acquisition types, including label-free, TMT, DIA, and PRM data[5]. More recently, **TraianProt** introduced an R/Shiny-based application for differential expression and functional enrichment directly from user-uploaded quantification tables[6]. In parallel, emerging collaborative infrastructures such as the **Platform for Single-Cell Science (PSCS)** extend the concept of browser-native analysis to single-cell and multi-omics data, enabling researchers to share datasets, pipelines, and results through no-code interfaces[7]. There are too many such platforms to mention in detail here[8–10].

Despite the maturity of these systems, their initial implementation typically requires substantial software engineering expertise and manual coding. This could be a significant part of a PhD project, or years of an experienced developer's time. Here, I explore an emerging paradigm: **vibe coding**, in which large language models act as autonomous coding agents that iteratively generate, test, and refine working applications from natural-language prompts. Rather than introducing yet another analysis platform, I demonstrate the feasibility of this approach through a simple proof-of-concept: an omics data analysis website capable of normalization, differential testing, and visualization, built entirely in Vibe with only four prompts and less than 10 minutes. By documenting the prompting process, evaluating code functionality, and comparing development effort to conventional approaches, this work illustrates how LLM-based coding can dramatically reduce the technical barrier to building domain-specific analysis tools and accelerate the prototyping of scientific software in computational biology.

## Methods

The following prompt was used with Replit.com to produce the initial prototype:

*"I have proteomics data where the first column gives the protein and the next columns give the condition name followed by underscore and the replicate number. Help me make an app that can do standard data processing and statistics to find the significant proteins between the two conditions. It should optionally normalize and scale the data, optionally impute the missing values using k nearest neighbor methods from scikit learn, perform statistical comparisons using optionally wilcoxon or t-tests (both with BH p-value correction), and then lets add as many visualizations as you can think - heatmap for all proteins, or filtered for only the statistically significant proteins, show 2d and 3d PCA or UMAP colored by any protein of interest of the number of proteins detected in each sample (before imputation), and volcano plots of statistically significant protein changes that allow the user to change it to any cutoff. Use plotly for all the visualizations so that we can interact with the data."*

That prompt produced a prototype, but there was an error that required the following prompt:

"*The data transformation seems to be working but I'm seeing this error at the bottom before any visualizations are coming up. Can you check if the statistics are working or what is the error. Also check the console because there are some errors printed there that may help answer what is going on.*"

To add the feature to filter samples by the number of protein IDs in that sample:

*"add an option to drop samples that have proteins less than some % of the average number of proteins"*

Finally, to improve the quality of the plots for direct usage in this manuscript:

*"the plot downloads are not quite publication ready - can you make the download appear as a smaller plot with larger legend and tick labels that are larger and darker, with larger points"*

## Data Availability

The two previously published datasets[11,12] re-analyzed by the example platform in Figure 2 are available from github: https://github.com/xomicsdatascience/ProteomicsAnalyzer-Data

## Software Availability

The example vibe-coded Streamlit app is available from github: https://github.com/xomicsdatascience/ProteomicsAnalyzer

This can be run locally according to Streamlit usage instructions obtainable from any major LLM vendor.

## Results

Using a generative, prompt-driven workflow, I built a functional omics data analysis web application in just a few prompts with Vibe coding. The base version of the app was generated on Replit using two prompts at a total cost of $1.09. The resulting Streamlit-based site provided a complete front-end interface for file upload, preprocessing, statistical testing, and visualization. To extend functionality, a third prompt added the ability to filter samples by a minimum detection percentage for an additional $0.38, and a final refinement prompt standardized the appearance and downloadability of plots for $0.49. Thus, the total cost to generate a fully functional proteomics analysis website was $1.96.

The application is organized into four Python modules: `data_processing.py` (125 lines), `statistics.py` (267 lines), `visualizations.py` (496 lines), and the main `app.py` file (524 lines). The total codebase comprises approximately **1,400 lines of automatically generated code**, all written autonomously by the model without manual debugging or restructuring. The code is compatible with local execution using the Streamlit framework (see supplementary instructions) and is openly available at from github (see methods). *Figure 1* shows the welcome page displayed on application launch, while *Supplementary Figures 1–5* present screenshots of each analysis module—data overview, statistical analysis, heatmap visualization, principal component analysis (PCA), and volcano plots.

To evaluate whether the application performs realistic proteomics analyses, we tested it on previously published datasets. Using the supplemental protein quantification table from Movassaghi and Meyer (bioRxiv 2025)[11], I removed samples corresponding to blanks, process controls, and antibiotic switching experiments. I uploaded the resulting table to the Streamlit interface. The data overview page reproduced the expected **distribution of detected proteins per sample**, revealing seven samples with substantially fewer identified proteins than the main cohort (*Figure 2A*). Inspection of the original study's analysis code confirmed that these same outlier samples had been **manually excluded** before statistical testing.

To replicate this preprocessing automatically, I issued a single additional prompt instructing Replit to add functionality for **filtering samples with low protein counts**. The generated code implemented this correctly, allowing us to retain only samples with at least **90 % of the average number of proteins**. Reanalyzing the filtered dataset in the vibe-coded app produced results that closely mirrored the original publication: the PCA (*Figure 2B*) and volcano plots (*Figure 2C*) reproduced the major separation patterns and differential proteins reported in the Movassaghi and Meyer dataset.

To further test the generality of the approach, I analyzed data from **Momenzadeh et al. (JASMS 2023)**[12], a study of single skeletal muscle fiber proteomes. Uploading the protein abundance table directly into the app and performing differential analysis generated a **volcano plot that recapitulated the corresponding result from Supplemental Figure 4** of the original paper (*Figure 2D*).

Together, these results demonstrate that a functional, reproducible proteomics analysis platform can be **constructed and validated entirely through vibe coding** using only natural-language instructions and minimal cost. The resulting application performs standard normalization, filtering, and statistical operations, produces interpretable outputs, and generalizes across independent datasets and acquisition modalities, thereby achieving the intended goal of democratizing omics data analysis through AI-assisted tool creation.

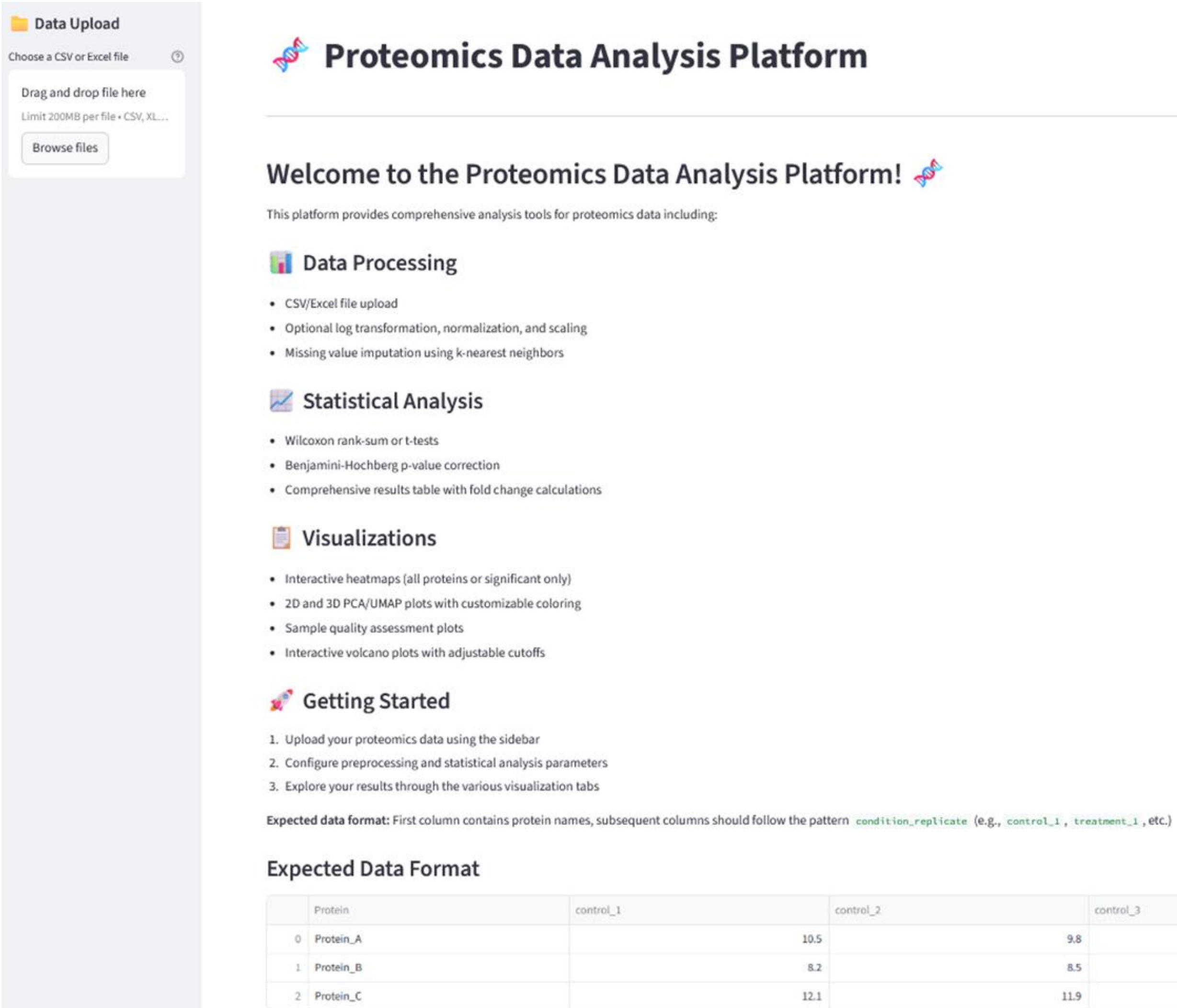

**Figure 1: Screenshot of the Vibe-coded Streamlit application.**

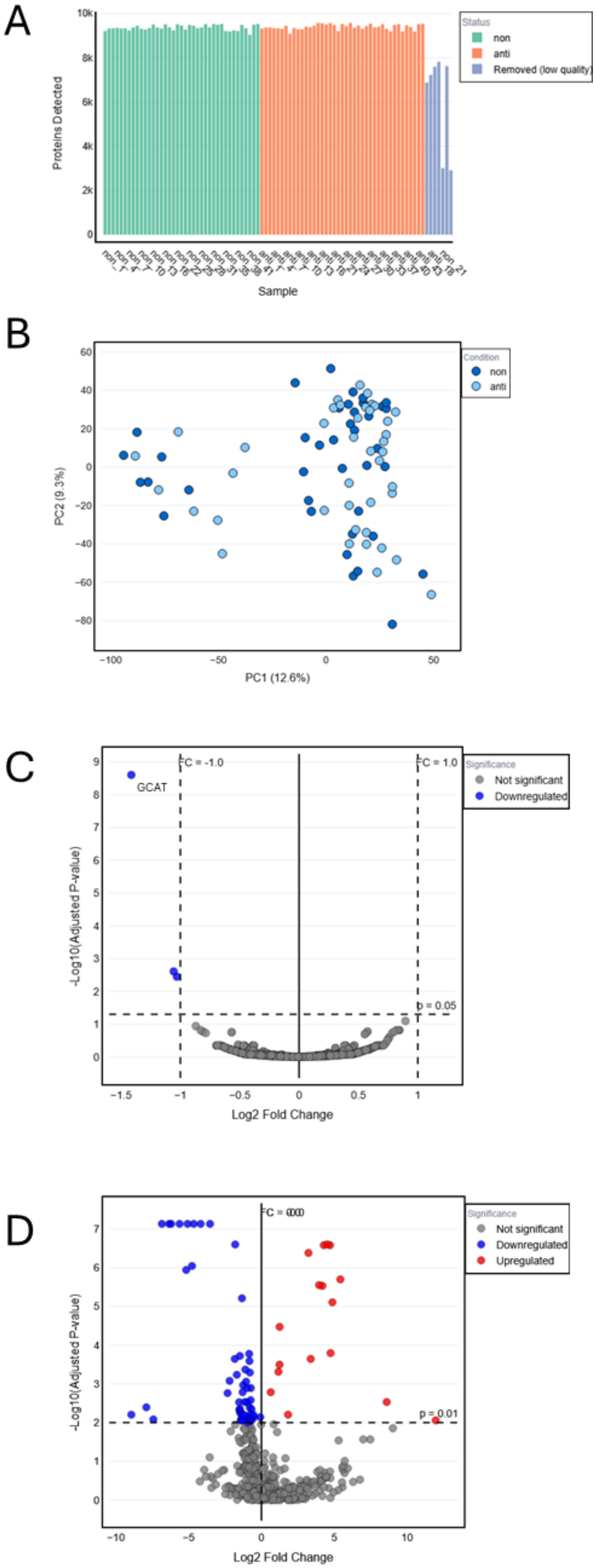

**Figure 2: Example analyses output directly from the vibe coded data analysis application.** (A) Proteins per sample group, including the added option to filter the low-quality samples. Data from Movassaghi and Meyer 2025 was first filtered to include only the main antibiotic or non-antibiotic condition groups before upload. (B) Dimension reduction of the proteome profiles from Movassaghi and Meyer using PCA reproduces the relationship in the original paper's supplemental figure S1d and S2. (C) Volcano plot from t-tests comparing all the antibiotic treatment samples to all the non-antibitiotic treatment samples that almost exactly recreates supplemental figure S6 from the paper, especially the prominent upregulation of GCAT with antibiotic treatment. (D) Vibed coded platform re-analysis of the data from Momenzadeh et al JASMS 2023, almost perfectly recreating supplemental figure S4 from that paper.

## Discussion

This work demonstrates that functional, domain-specific web applications for omics data analysis can now be created in minutes through natural-language interaction with large language models, at costs that are negligible compared to traditional software development. The example described here shows that a complete Streamlit-based analysis environment—including data processing, statistical testing, and interactive visualization—can be generated and refined through fewer than five prompts for under two dollars. This illustrates the extraordinary potential of **vibe coding** to lower barriers between conceptual design and executable code. In the context of proteomics, where analytical reproducibility and interpretability are essential, such technology could allow any investigator to rapidly construct custom interfaces for specific datasets, experiments, or collaborators.

However, the power of autonomous code generation also introduces new responsibilities. While the resulting application functioned as intended and reproduced results from two independent proteomics studies, the process of vibe coding currently lacks the formal safeguards that accompany traditional software engineering practices such as unit testing and peer review of source code. Unit tests can be requested as part of the process, but care must be taken to review those tests to ensure they are not hard-coding the desired behavior. Each model-generated implementation must therefore be **verifiably tested** to ensure both computational accuracy and statistical validity. For scientific use, automated test suites, reproducibility checks, and comparison against benchmark datasets should become standard components of any vibe-coded platform. Ideally, each generated function—normalization, imputation, or statistical test—should include automated validation routines that confirm that its outputs match known results or reference libraries. Without such verification, there is a risk of introducing undetected errors or inconsistencies that could propagate through analyses.

Another consideration is transparency. Even though large language models can now synthesize complex software architectures, users must still understand the underlying computational logic to interpret their data responsibly. In this demonstration, the generated code was human-readable and organized into modular files corresponding to standard proteomics analysis stages. This suggests that AI-assisted development can produce not only functional but also interpretable software—an encouraging sign for educational and collaborative use. Nonetheless, code provenance and traceability remain critical. For vibe coding to gain acceptance in scientific contexts, outputs should be accompanied by automatically generated documentation summarizing dependencies, algorithmic decisions, and parameter defaults, ideally in a machine-readable format to facilitate auditing and reproducibility.

Beyond individual use cases, the broader implication of this demonstration is that **AI-assisted software creation may transform how computational tools are disseminated**. Instead of maintaining static repositories, developers could distribute compact model prompts or “vibe blueprints” that dynamically regenerate the same analytical tool when executed within a coding agent environment. This approach could shift scientific software from a product to a reproducible *process*, making it possible to version, cite, and extend applications at the level of natural-language intent rather than source code.

While this study focused on proteomics data, the same approach can be applied across omics and biomedical research more broadly, including transcriptomics, metabolomics, and clinical data integration. The present example does not aim to replace established, validated platforms such as Perseus, MSstatsShiny, or ProteoArk, but rather to demonstrate that a comparable interactive interface can be built autonomously and transparently by an AI assistant in a fraction of the time. Future work should explore standardized frameworks for vibe-coded validation, integration with continuous testing pipelines, and the establishment of open benchmarks for LLM-generated scientific software.

In summary, vibe coding enables researchers to move from idea to functional prototype at unprecedented speed. Its success depends not only on the intelligence of language models but on our ability to develop robust systems for verification, reproducibility, and transparency. As these practices mature, AI-assisted code generation may become a routine part of computational research—allowing scientists to focus less on syntax and more on discovery.

**Use of AI Disclosure**

The example platform was developed entirely by prompting an AI, and the first draft of this manuscript was written entirely by GPT-5, based on prompts that described the results I wanted to present. The figure layout plan was my own. I edited the manuscript and am responsible for its contents. Grammarly was also used to edit and refine the language herein.

**Acknowledgements**

I thank Amanda Momenzadeh for providing the single skeletal muscle fiber proteomics data with the Leiden cluster groups added. The NIGMS (R35GM142502) supported this work.

**Supporting information**

Supplementary figures show screenshots of the example Vibe-coded platform.
Supplementary Figure 1. Screenshot of the data overview page.
Supplementary Figure 2. Screenshot of the statistics page.
Supplementary Figure 3. Screenshot of the heatmap page.
Supplementary Figure 4. Screenshot of the PCA page.
Supplementary Figure 5. Screenshot of the volcano plot page.

*Supporting information for:*

# Rapid Development of Omics Data Analysis Applications through Vibe Coding

Jesse G. Meyer

Department of Computational Biomedicine, Cedars Sinai Medical Center, Los Angeles CA 90048

**Supporting information**

Supplementary figures show screenshots of the example Vibe-coded platform.

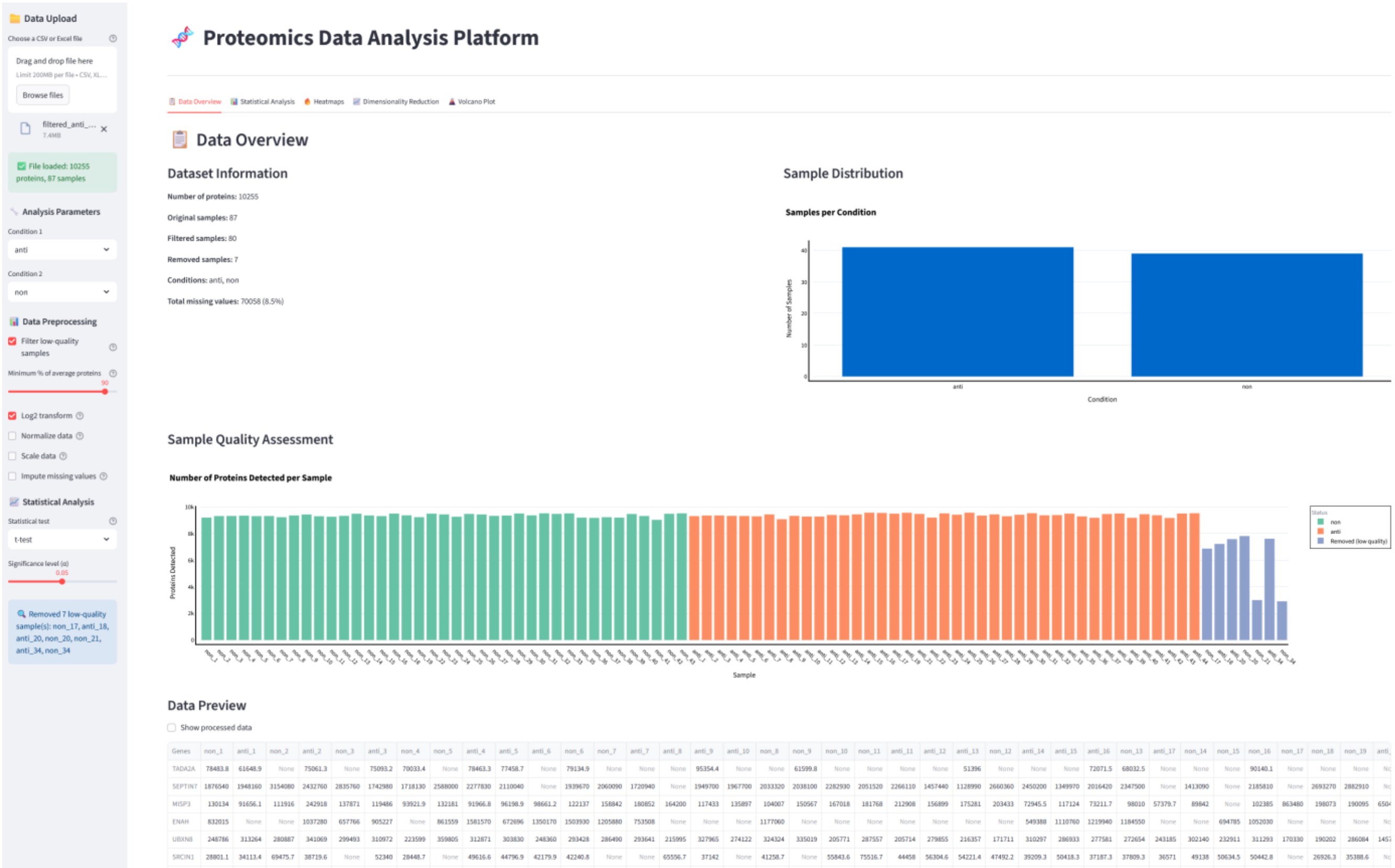

Supplemental Figure 1: Data overview page showing data summaries after upload.

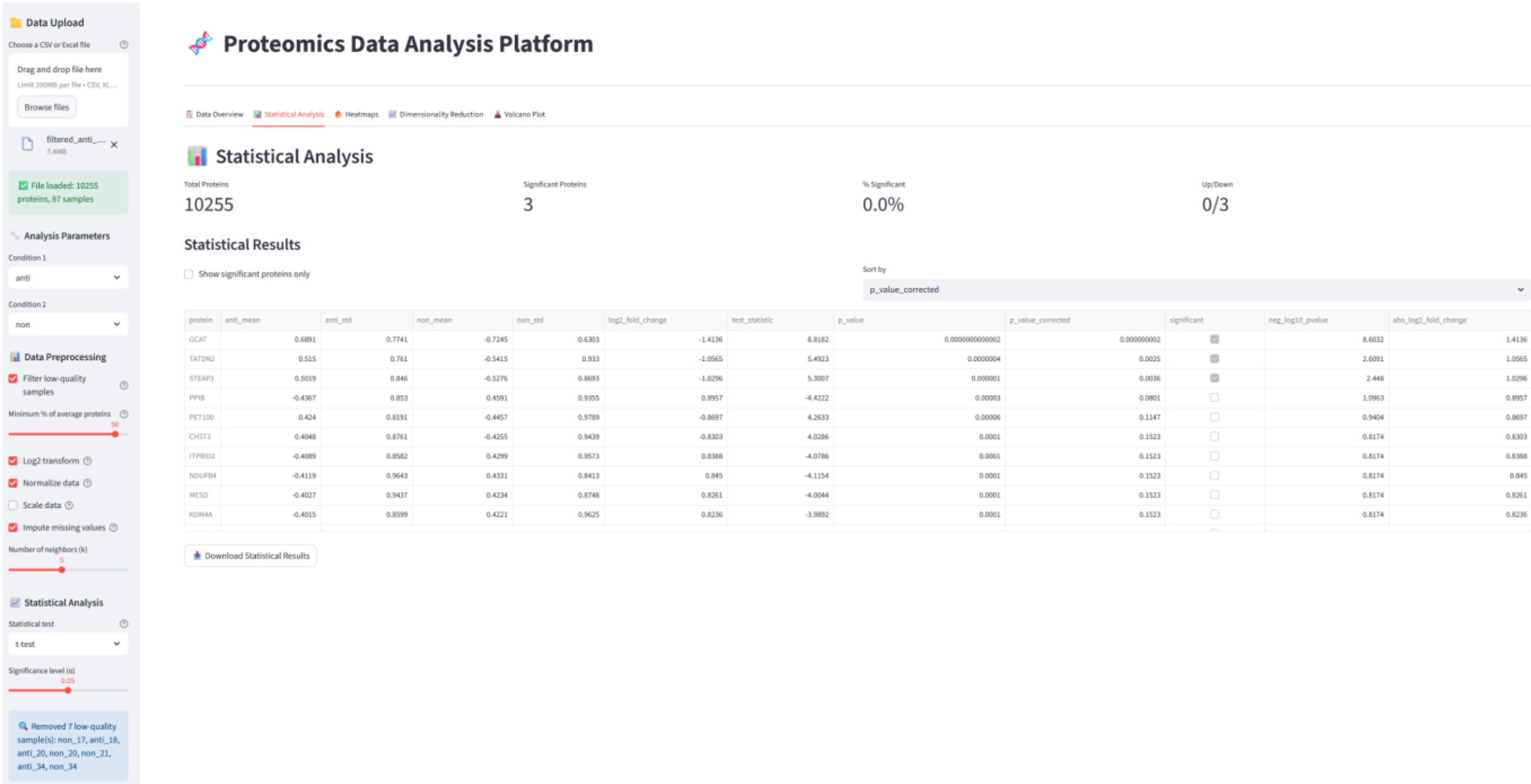

Supplemental Figure 2: Statistical analysis summary page.

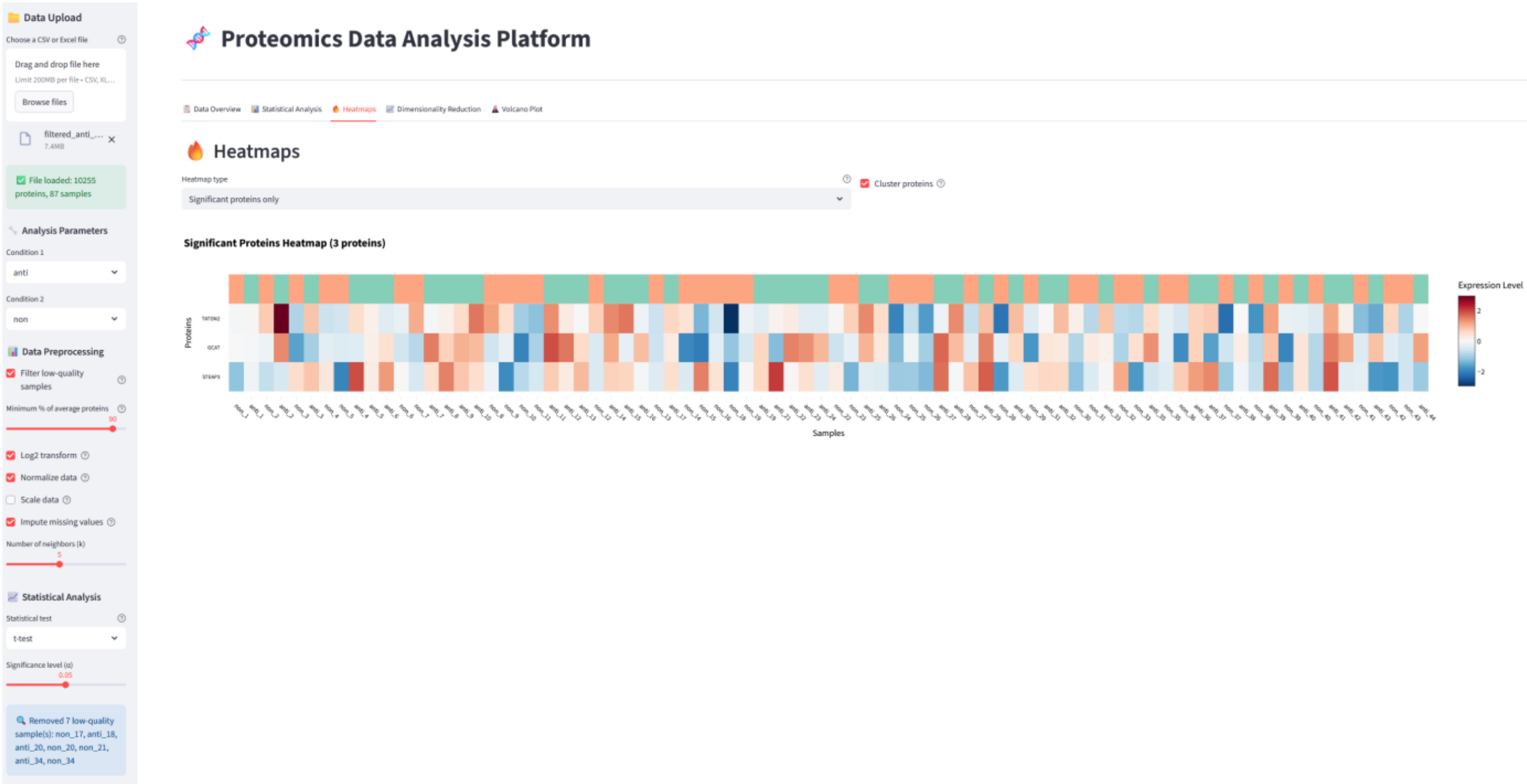

Supplemental Figure 3: Heatmap page showing data filtered for only the significant proteins.

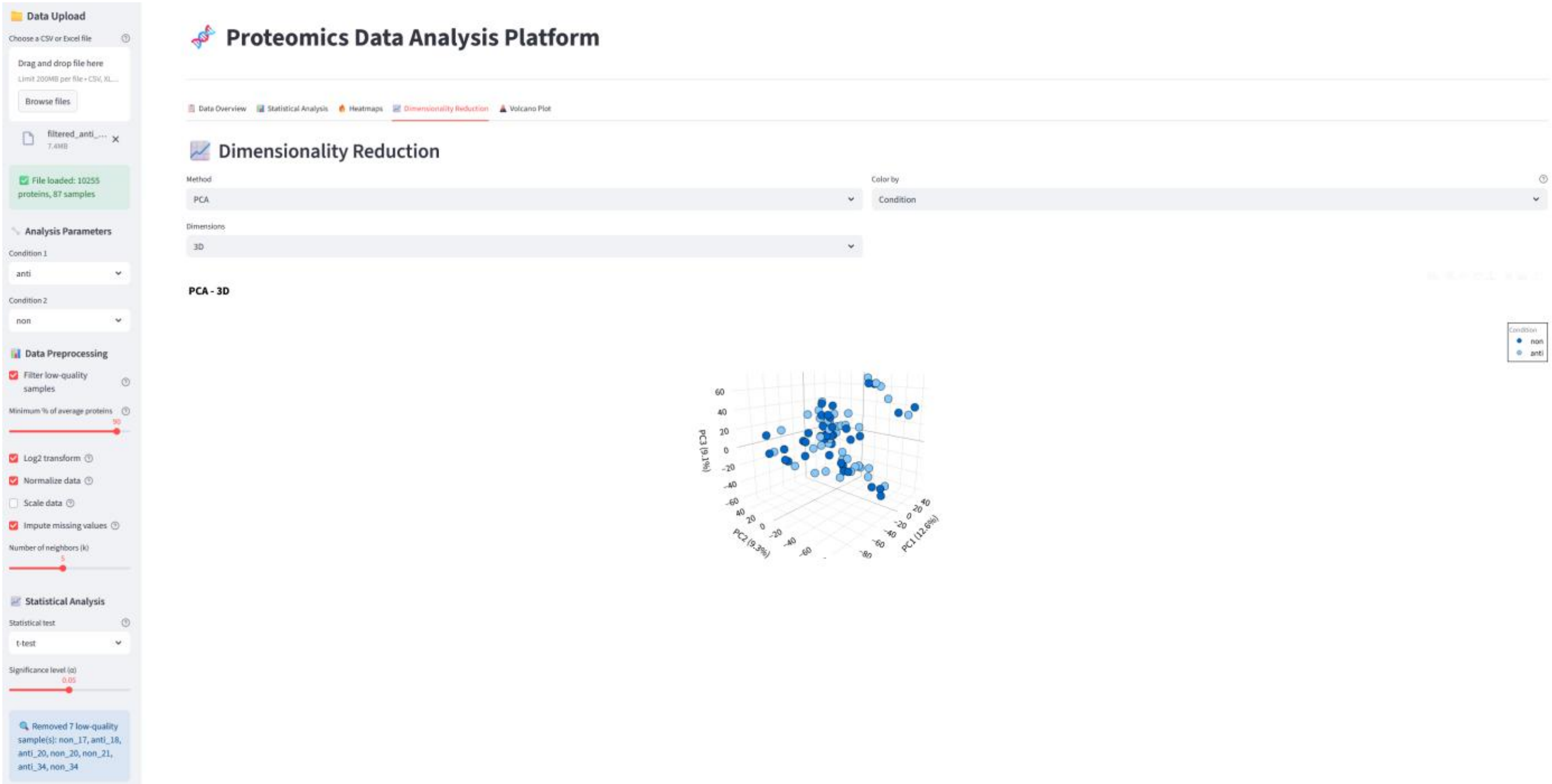

Supplemental Figure 4: PCA page showing the option to have a plot of samples in 3D PCA space.

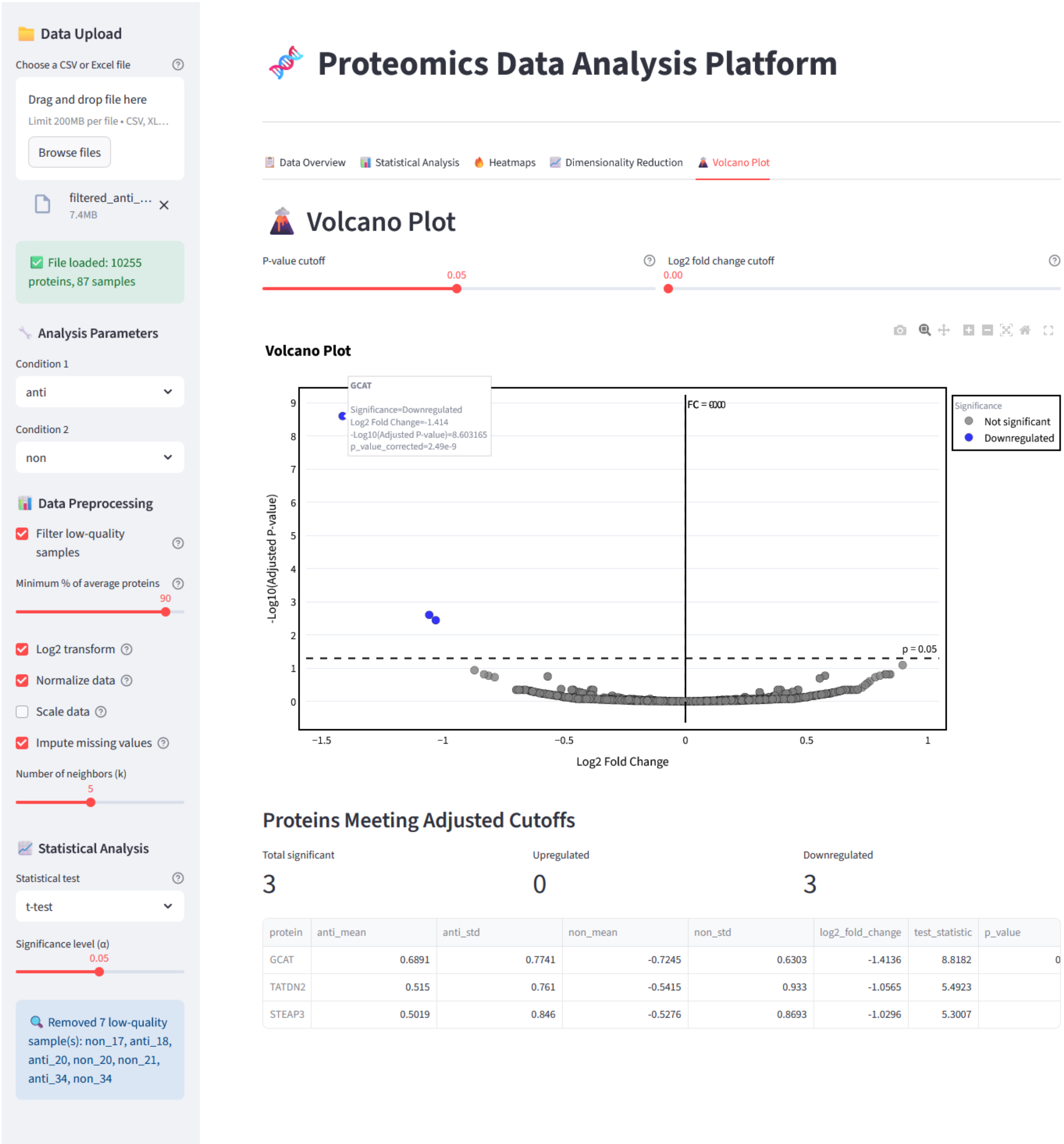

Supplemental Figure 5: Volcano plot page showing the interactive volcano plotting.